\documentclass[twocolumn,preprintnumbers,secnumarabic,amsmath,amssymb,superscriptaddress,nofootinbib]{revtex4}

\usepackage{graphicx}
\usepackage{dcolumn}
\usepackage{bm}

\def\beq{\begin{equation}}
\def\eeq{\end{equation}}
\def\be{\begin{equation}}
\def\ee{\end{equation}}
\def\bea{\begin{eqnarray}}
\def\eea{\end{eqnarray}}
\def\d{{\rm d}}

\begin{document}

\preprint{ITEP-TH/12-07}
\preprint{PUPT-2235}

\title{A Delicate Universe}

\medskip\
\author{Daniel Baumann}
\affiliation{%
Department of Physics, Princeton University, Princeton, NJ 08544}

\author{Anatoly Dymarsky}
\affiliation{%
Department of Physics, Princeton University, Princeton, NJ 08544}

\author{Igor R. Klebanov}
\affiliation{%
Department of Physics, Princeton University, Princeton, NJ 08544}
\affiliation{%
Princeton Center for Theoretical Physics, Princeton University, Princeton, NJ 08544}

\author{Liam McAllister}%
\affiliation{%
Department of Physics, Princeton University, Princeton, NJ 08544}

\author{Paul J. Steinhardt}
\affiliation{%
Department of Physics, Princeton University, Princeton, NJ 08544}
\affiliation{%
Princeton Center for Theoretical Physics, Princeton University, Princeton, NJ 08544}

\date{\today}
\begin{abstract}
We investigate whether explicit models of warped D-brane inflation are possible in string compactifications.
To this end, we study the potential for D3-brane motion in a warped conifold that includes holomorphically-embedded D7-branes involved in moduli stabilization. The presence of the D7-branes significantly modifies the inflaton potential.
We construct an example based on a very simple and symmetric embedding due to Kuperstein, $z_1 = $ constant,
in which it is possible to fine-tune the potential so that slow roll inflation can occur.
The resulting model is rather delicate: inflation occurs in the vicinity of an inflection point, and the cosmological predictions are extremely sensitive to the precise shape of the potential.

\end{abstract}

\maketitle

{\bf Introduction.}\
String theory is a promising candidate for the theoretical underpinning of the inflationary paradigm \cite{Inflation}, but explicit and controllable models of inflation in string theory have remained elusive.  In this Letter we ask whether explicit working models are possible in the setting of slow roll warped D-brane inflation \cite{Dvali,KKLMMT}, in which the inflaton field is identified with the location of a mobile D3-brane in a
warped throat region \cite{KS} of the compactification manifold.  As explained in \cite{KKLMMT}, moduli stabilization introduces potentially fatal corrections to the inflaton potential in this scenario.  Some of these corrections arise from complicated properties of the compactification \cite{BHK} and have been computed only recently \cite{BDKMMM}.

The attitude taken in most of the literature on the subject ({\it{cf.}}~\cite{KKLMMT,LindeReview}) is that because of the vast number and complexity of string vacua, in some nonzero fraction of them it should be the case that the different corrections to the inflaton potential cancel to high precision, leaving a suitable inflationary model.  This expectation or hope has never been rigorously justified (but see \cite{Sandip} for a promising proposal), and there is no guarantee that the correction terms can ever cancel: for example, it may be the case that the correction terms invariably have the same sign, so that no cancellation can occur.
In this Letter we report the results of a systematic investigation into whether or not this hope of fine-tuned cancellation can in fact be realized.  Further details will appear in \cite{LongPaper}.  The new ingredient that makes this work possible is the result of \cite{BDKMMM} for a correction to the volume-stabilizing nonperturbative superpotential.  As explained in \cite{BHK,GM,BDKMMM}, this effect corresponds to the interaction between the inflationary D3-brane and the moduli-stabilizing wrapped branes \cite{KKLT}, {\it{i.e.}} D7-branes or Euclidean D3-branes wrapping a four-cycle of the compact space.  The location of these wrapped branes therefore becomes a crucial parameter in the D3-brane potential.

In a recent paper \cite{Burgess}, Burgess {\it et al.} showed that for a particular embedding of the D7-branes, the Ouyang embedding \cite{Ouyang}, the correction to the inflaton potential from the term computed in \cite{BDKMMM} vanishes identically.  In this case the potential is always too steep for inflation, independent of fine-tuning.  We have found a similar problem \cite{LongPaper} in the large class of D7-brane embeddings described in \cite{Arean}.  Here we consider a more promising case, a simple holomorphic embedding due to Kuperstein \cite{Kuperstein}.
For fine-tuned values of the microphysical parameters, the potential for radial motion of a D3-brane in this background contains an approximate inflection point around which slow roll inflation can occur.  This potential is not of the form anticipated by previous authors: the D7-branes have no effect on the quadratic term in the inflaton potential, but instead cause the potential to flatten in a small region far from the tip of the conifold.  We emphasize that arranging for this inflection point to occur inside the throat region, where the metric is known and our construction is self-consistent, imposes a severe constraint on the compactification parameters.  Moreover, inflation only occurs for a bounded range about the inflection point, which requires a high degree of control over the initial conditions of the inflaton field.

We employ natural units where  $M_P^{-2} = 8\pi G \equiv 1$.

\begin{figure}[htbp!]
    \centering
        \includegraphics[width=0.40\textwidth]{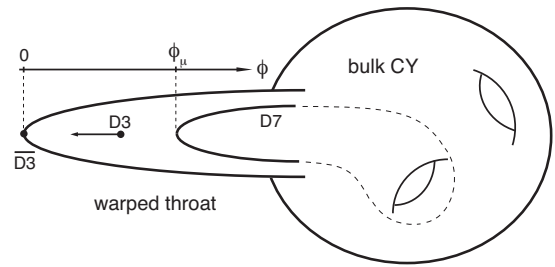}
    \caption{Cartoon of an embedded stack of D7-branes wrapping a four-cycle, and a mobile D3-brane,
in a warped throat region of a compact Calabi-Yau.}
    \label{fig:throat}
\end{figure}

{\bf The Compactification.}\
Our setting is a flux compactification \cite{GKP,FluxReview} of type IIB
string theory on an orientifold of a Calabi-Yau threefold, or, more generally, an F-theory compactification.
We suppose that the fluxes are chosen so that the internal space has a warped
throat region, and that $n>1$ D7-branes supersymmetrically wrap a four-cycle that extends
into this region (see Figure \ref{fig:throat}).  As a concrete example of this local geometry,
we consider the warped version \cite{KS} of the deformed conifold $\sum z_i^2 = \varepsilon^2$,
where $z_i$ are coordinates on $\mathbb{C}^4$.  Assuming that the D3-brane is far from the tip of the conifold,
we may neglect the deformation $\varepsilon$.
We choose $z_\alpha=(z_1,z_2,z_3)$
as the three independent complex D3-brane coordinates, and use the conifold constraint
to express $z_4$ in terms of them.
We suppose that this throat is glued into a compact space, as in \cite{GKP}, and for simplicity we take
this space to have a single K\"ahler modulus $\rho$.  Moduli stabilization
\cite{KKLT} relies on the fact that strong gauge dynamics on suitable D7-branes generates
a nonperturbative superpotential, $W_{\rm np} = A(z_\alpha) \exp[{-a \rho}]$,
where $a=\frac{2\pi}{n}$.
The D7-brane embedding is specified by a single holomorphic equation, $ f(z_\alpha) = 0 $,
and the result of \cite{BDKMMM} is that
\beq \label{equ:BDKMMMResult}  A(z_\alpha) = A_0 \left(\frac{f(z_\alpha)}{f(0)} \right)^{1/n} \, , \eeq
where $A_0$ is independent of the D3-brane position $z_{\alpha}$.
Including the flux superpotential \cite{GVW} $W_{\rm flux} = \int G\wedge \Omega \equiv W_0$, the total superpotential is
$W = W_0 + A(z_\alpha) \exp[{-a\rho}] $.  Next, the DeWolfe-Giddings K\"ahler potential \cite{DeWG} is
\beq \label{equ:KKK}
{\cal K}(\rho,\bar \rho ,z_\alpha, \bar z_\alpha)  = -
3 \log [\rho + \bar \rho - \gamma k]  \equiv - 3 \log U\, ,
\eeq
where $k(z_\alpha, \bar z_\alpha)$ is the K\"ahler potential of the Calabi-Yau space,
and $\gamma$ is a constant \cite{LongPaper}.
Well inside the throat but far from the tip, we may use the K\"ahler potential
of the conifold \cite{Candelas},
\beq \label{equ:randz}
k = \frac{3}{2} \left(
\sum_{i=1}^4 |z_i|^2 \right)^{2/3} = \frac{3}{2} r^2 \, . \eeq
Then the F-term potential is \cite{Burgess}, \cite{LongPaper}
\bea \label{equ:Fterm} V_F &=&
\frac{1}{3 U^2} \Biggl[ (\rho + \bar \rho)
|W_{,\rho}|^2 - 3 (\overline{W} W_{, \rho} + c.c.) \nonumber \\&& +
\frac{3}{2} (\overline{W_{,\rho}} z^\alpha W_{,\alpha} + c.c.) +
\frac{1}{\gamma} k^{\alpha \bar \beta} W_{,\alpha}
\overline{W_{,\beta}} \Biggr]\, , \eea
where
\beq
k^{\alpha \bar \beta}=
r \bigg [
\delta^{\alpha \bar \beta}+ \frac{1}{2}
{z_\alpha \bar z_\beta \over  r^3}  - {z_\beta \bar z_\alpha \over r^3}
\bigg ]\, .
\eeq
To this we add the contribution of an anti-D3-brane at the tip of the deformed conifold \cite{KKLMMT},
\beq  V_D = D(r)U^{-2} \, , \qquad  D(r) \equiv D \left(1- \frac{3D}{16\pi^2}\frac{1}{(T_3 r^2)^2} \right) \, , \eeq where
$D= 2T_3/h_0$, $T_3$ is the D3-brane tension, and $h_0$ is the warp factor \cite{KS} at the tip. \\

{\bf Towards Fine-Tuned Inflation.}\
To derive the effective single-field potential, we consider radial trajectories that are
stable in the angular directions, so that the dynamics of the angular fields becomes trivial.
We also integrate out the massive volume modulus, incorporating the crucial fact that the
volume shifts as the D3-brane moves \cite{LongPaper}.  Then the canonically-normalized inflaton $\phi \equiv r \sqrt{\frac{3}{2} T_3}$
parameterizes the motion along the radial direction of the throat.  To investigate the possibility of sustained inflation, we consider
the slow-roll parameter $\eta = V''/V$.  We find $\eta = \frac{2}{3} + \Delta \eta(\phi)$,
where $\Delta \eta$ arises from the dependence (\ref{equ:BDKMMMResult}) of the superpotential on $\phi$.  Slow-roll inflation is possible near $\phi=\phi_0$ if $\Delta \eta(\phi_0) \approx -\frac{2}{3}$. Here, using the explicit result of \cite{BDKMMM} for $A(\phi)$, we compute $\Delta \eta$ and determine whether the full potential can be flat enough for inflation.\footnote{
For the special case of the Ouyang embedding, $z_1 + i z_2 = \mu$, Burgess {\it et al.} proved a simple no-go result for fine-tuned brane inflation \cite{Burgess}.
They found that for this particular example, $\Delta \eta$ vanishes along the angularly stable trajectory.
We have found similar `delta-flat' trajectories \cite{LongPaper}
for all embeddings in the infinite class studied in \cite{Arean}.
These trajectories cannot support slow roll inflation, no matter how the parameters of the potential are tuned. In this paper, we study an embedding for which there is no delta-flat direction.}

A reasonable expectation implicit in prior work on the subject is that there exist fine-tuned values of the microphysical parameters for which $\Delta \eta(\phi) \approx -\frac{2}{3} $, {\it{i.e.}} the correction to the inflaton potential arising from $A(\phi)$ includes a term quadratic in $\phi$, which, for a fine-tuned value of its coefficient, causes $\eta$ to be small for a considerable range of $\phi$.  However, we make the important observation that the functional form of (\ref{equ:BDKMMMResult}) implies that there is actually {\it no} purely quadratic correction.  To see this we note that $A$ is a holomorphic function of the $z_\alpha$ coordinates, which, by (\ref{equ:randz}), scale with radius as $z_{\alpha} \propto \phi^{3/2}$.  Thus, the presence of $A(\phi)$ in the form (\ref{equ:BDKMMMResult}) does
not lead to new quadratic terms in (\ref{equ:Fterm}).  This is concrete evidence against the hope of a fine-tuned cancellation of the inflaton mass over an extended range of $\phi$.

However, as we now explain, there exists a simple example in which a different sort of
cancellation can occur.
Kuperstein \cite{Kuperstein} studied the D7-brane embedding $z_1=\mu$, where we may assume that
$\mu \in \mathbb{R}^{+}$. This embedding, and the potential in this background, preserve an $SO(3)$
subgroup of the $SO(4)$ global symmetry acting on the $z_i$ coordinates of the deformed conifold.
To find a purely radial trajectory that is stable in the angular directions, we consider the variation $\delta z_1$ while keeping the radius $r$ fixed.
We then require the first variation of the potential $\delta V = V(z_1+\delta z_1,r,\rho)- V(z_1,r, \rho)$ to vanish for all $r$,
and the second variation $\delta^2V$ to be non-negative.
The extremality constraint $\delta V=0$ specifies the radial trajectories $z_1=
\pm \frac{1}{\sqrt{2}} r^{3/2}$, $z_2 = \pm i z_1$.
A detailed study \cite{LongPaper} of the angular mass matrix $\delta^2 V$ reveals that the trajectory along $z_1 = + \frac{1}{\sqrt{2}} r^{3/2}$ is unstable, while the trajectory along $z_1 = - \frac{1}{\sqrt{2}} r^{3/2}$ is stable in all angular directions.
After integrating out the imaginary part of the K\"ahler modulus $\rho$, which amounts to the replacement $A \to |A|$ \cite{LongPaper},
the potential along the latter trajectory is given in terms of the radius $r$ (or the canonical inflaton $\phi$) and the real-valued volume modulus $\sigma \equiv \frac{1}{2} (\rho + \bar \rho)$, as
 \cite{LongPaper}
\bea
\label{equ:VTwoField}
V(\phi,\sigma)  &=& \frac{a |A_0|^2}{3}\frac{e^{-2a\sigma} }{U^2(\phi,\sigma)} g(\phi)^{2/n} \Biggl[
2 a \sigma +6 \nonumber \\
&& -\ 6 e^{a \sigma}  \frac{|W_0|}{|A_0|} \frac{1}{g(\phi)^{1/n}}
+ \frac{3 c }{n} \frac{\phi}{\phi_\mu} \frac{1}{g(\phi)^{2}}  \nonumber \\
&& -  \frac{3}{n} \frac{1}{g(\phi)}  \frac{\phi^{3/2}}{\phi_\mu^{3/2}}  \Biggr]  + \frac{D(\phi)}{U^2(\phi,\sigma)}\, .
\eea
Here $g(\phi) \equiv \frac{f(\phi)}{f(0)} =  1 + \bigl(\frac{\phi}{\phi_\mu} \bigr)^{3/2}$, and
$\phi_\mu^2 \equiv \frac{3}{2} T_3
(2\mu^{2})^{2/3}$ denotes the minimal radial location of the
D7-branes.
We have also introduced $ c^{-1} \equiv 4\pi\gamma(2\mu^2)^{2/3}$, used \cite{LongPaper} $\gamma = \sigma_0 T_3/3 $,
and defined $U(\phi,\sigma)  \equiv 2 \sigma - \frac{\sigma_0 }{3} \phi^2$. The parameter $\sigma_0$ is the stabilized value of the K\"ahler modulus in the absence of the D3-brane (or when the D3-brane is near the bottom of the throat).
Now, for each value of $\phi$ we carry out a constrained minimization of the potential,
{\it i.e.}~we find
$\sigma_\star(\phi)$
such that $\left.\frac{\partial V}{\partial \sigma}\right|_{\sigma_{\star}(\phi)} = 0$.
The function $\sigma_\star(\phi)$
 may either be computed numerically or fitted to high accuracy by the approximate expression \cite{LongPaper} \beq
\sigma_\star(\phi) \approx \sigma_0 \left[ 1 +
\frac{1}{n\,  a \sigma_0} \left( 1 - \frac{1}{2
a\sigma_0}\right) \left({\phi\over \phi_\mu}\right)^{3/2} \right]\, . \eeq
Substituting $\sigma_\star(\phi)$ into (\ref{equ:VTwoField}) gives our main result, the effective single-field potential
$\mathbb{V}(\phi) \equiv V(\phi, \sigma_\star(\phi))$.

\begin{figure}[h!]
    \centering
        \includegraphics[width=0.40\textwidth]{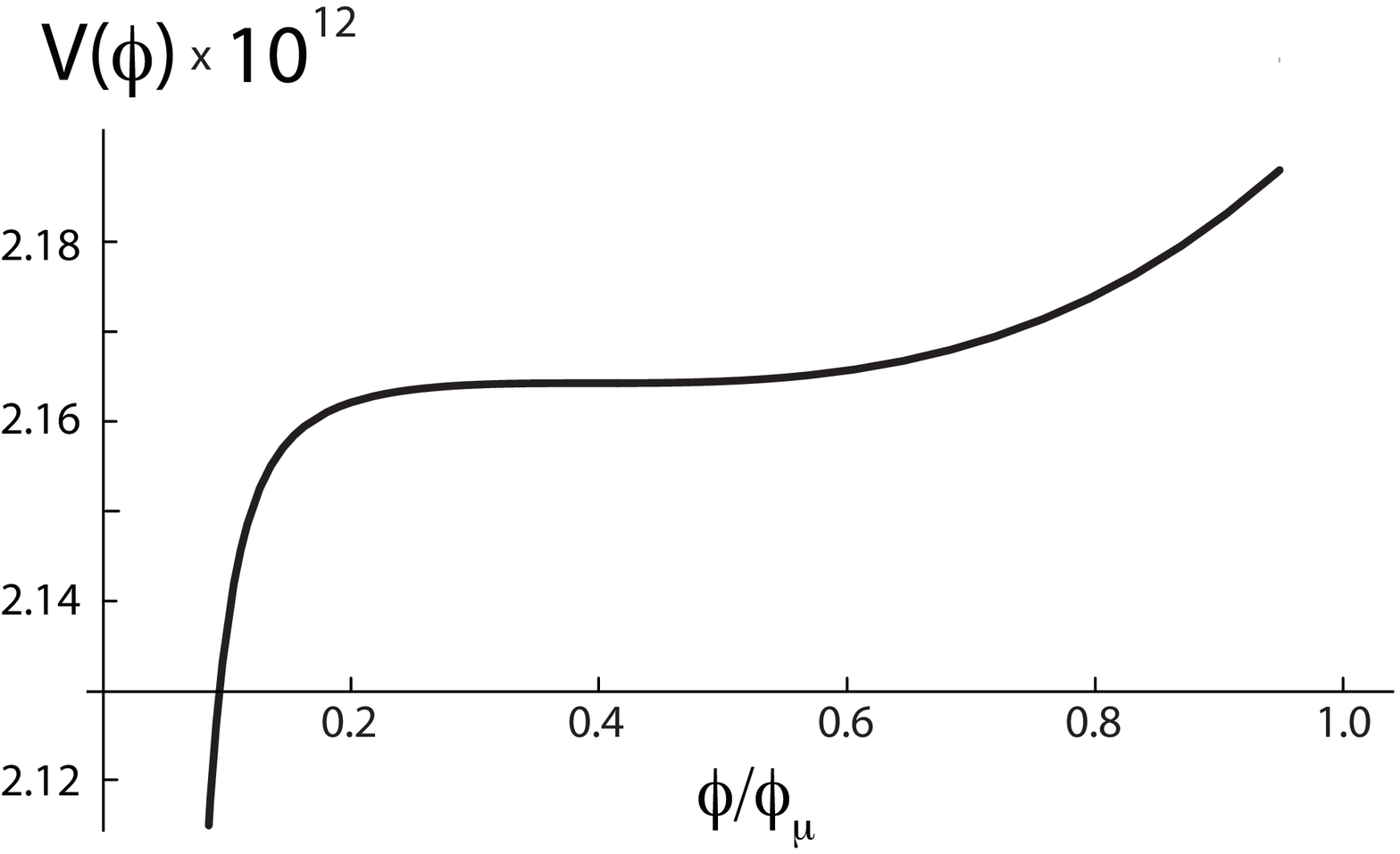}
    \caption{{\bf Inflaton potential $\mathbb{V}(\phi)$.}\\
    Compactification data: $n=8$, $\phi_\mu = \frac{1}{4}$, $A_0 = 1$,
   $W_0= - 3.432 \times 10^{-4}$,
   $D = 1.2 \times 10^{-8} $, which imply $a \sigma_0 \approx 10.1$.}
    \label{fig:pot}
\end{figure}

For generic values of the compactification parameters, $\mathbb{V}$ has a metastable minimum at some distance from the tip.  In fact, one can show that the potential has negative curvature near the tip and positive curvature far away, so that by continuity, $\eta$ vanishes at some intermediate value $\phi_0$.  Then, one can find fine-tuned values of the D7-brane position $\phi_{\mu}$ for which this minimum is lifted to become an inflection point (see Figure \ref{fig:pot}).  This transition from metastability to monotonicity guarantees that $\epsilon = \frac{1}{2}(V'/V)^2$ can be made extremely small, so that prolonged slow-roll inflation is possible.
In our scenario, then, the potential contains an approximate inflection point at $\phi = \phi_0$, where $\mathbb{V}$ is very well approximated by the cubic
\beq
\label{equ:cubic}
\mathbb{V} = V_0 + \lambda_1 (\phi - \phi_0) +  \frac{1}{3!} \lambda_3 (\phi-\phi_0)^3\, ,
\eeq
for some $V_0, \lambda_1, \lambda_3$.

The number of $e$--folds derived from the effective potential (\ref{equ:cubic}) is
\beq
\label{equ:Ne}
N_e(\phi) = \int_{\phi_{\rm end}}^\phi \frac{\d \phi}{\sqrt{2 \epsilon}} =
\sqrt{\frac{2 V_0^2}{ \lambda_1 \lambda_3}} \arctan \left.  \left( \frac{ V_0 \eta(\phi)}{\sqrt{2 \lambda_1 \lambda_3}} \right) \right|_{\phi_{\rm end}}^\phi\, .
\eeq
Since $\eta$ is small only for a limited range of inflaton values, the number of $e$--folds is large only when $\epsilon$ is very small. This forces these models to be of the small field type.  The scalar spectrum on scales accessible to cosmic microwave background (CMB) experiments can be red, scale-invariant, or blue, depending on how flat the potential is.  That is,
$n_s-1 = \left. ( 2\eta -6 \epsilon)\right|_{\phi_{\rm CMB}} \approx 2 \eta(\phi_{\rm CMB})$,
where $\phi_{\rm CMB}$ corresponds to the field value when observable scales exit the horizon during inflation, say between $e$-folds 55 and 60.
The sign of $\eta(\phi_{\rm CMB})$, and hence of $n_s - 1$, depends on where $\phi_{\rm CMB}$ is relative to the inflection point.  If inflation only lasts for the minimal number of $e$--folds to solve the horizon and flatness problems then the scalar spectrum is blue.  If the potential is made more flat, so that $\epsilon$ is smaller, inflation lasts longer, and $\phi_{\rm CMB}$ is reduced, the spectrum can be red.  This sensitivity to the details of the potential reduces the predictivity of the scenario.\\

{\bf Microscopic Constraints.}\ A crucial consistency requirement is that the inflationary region around $\phi_0$, and the location $\phi_{\mu}$ of the tip of the wrapped D7-branes, should fit well inside the throat, where the metric is known.  As shown in \cite{BM}, the range of $\phi$ in Planck units is geometrically limited,
\beq \label{equ:BM} \Delta \phi < \frac{2}{\sqrt{N}} \, , \eeq
where $N \gg 1$ is the background D3-brane charge of the throat.  When combined with the Lyth bound \cite{Lyth}, this yields a sharp upper bound on the tensor signal in these models \cite{BM}.  Here we find that this same bound actually poses an obstacle to inflation itself: for an explicit inflationary model with the Kuperstein embedding of D7-branes, $\phi_\mu$ and $\phi_0$ must obey  (\ref{equ:BM}).  Although one can find
examples \cite{LongPaper} in which this requirement is met,
this imposes significant restrictions on the compactification. In particular, $N$ cannot be too large, implying that corrections to the supergravity approximation could be significant. \\

{\bf Conclusions.}\
We have assessed the prospects for an explicit model of warped D-brane inflation by including the known dangerous corrections
to the inflaton potential.
In particular, we have studied whether the hope of fine-tuning superpotential corrections to the inflaton potential to reduce the slow roll parameter $\eta$ can be justified.
As shown in \cite{LongPaper}, for a large class \cite{Arean} of holomorphic embeddings of wrapped D7-branes there are trajectories where
the potential is too steep for inflation, with no possibility of fine-tuning to avoid this conclusion \cite{Burgess}.
For the Kuperstein embedding \cite{Kuperstein}, fine-tuning is possible in principle,
and inflation can occur in a small region near an inflection point of the potential.
The requirement (\ref{equ:BM}) that this inflection point lies well inside the throat
provides stringent constraints on the compactification. Detailed construction of compactifications
where such constraints are satisfied remains an open problem.

This study illustrates the care that must be taken in assessing the prospects for inflationary cosmology in string theory. It appeared that warped D-brane inflation involved many adjustable parameters, including the D7-brane embedding and other compactification data, and so it was reasonable to expect that many working examples would exist.  However, the compactification geometry constrains these microphysical parameters
so that there is much less freedom to adjust the shape of the potential than simple parameter counting would suggest.

The problem of constructing a fully explicit model of inflation in string theory remains important and challenging.  Diverse corrections to the potential that are negligible for many other purposes can be fatal for inflation, and one cannot reasonably claim success without understanding all these contributions.  We have made considerable progress towards this goal, but have not yet succeeded: a truly exhaustive search for further corrections to the inflaton potential remains necessary.\\

{\bf Acknowledgements.}\
We thank C. Burgess, J. Conlon, O. DeWolfe, J. Distler, S. Kachru,  J. Maldacena, A. Murugan, J. Polchinski, G. Shiu, E. Silverstein, H. Tye, and B. Underwood for helpful discussions. This research was supported in part by the National Science Foundation
under Grant No. PHY-0243680 and by the Department of Energy under grant DE-FG02-90ER40542. 
The research of A.D. was supported in part by Grant RFBR 07-02-00878, and Grant for Support of Scientific
Schools NSh-8004.2006.2.
Any opinions, findings, and conclusions or recommendations expressed in this material are those of the authors and do not necessarily reflect the views of these funding agencies.

\begingroup\raggedright\endgroup

\end{document}